\documentclass[]{spie}
\usepackage{amsmath,amsfonts,amssymb}
\usepackage{graphicx}
\usepackage[colorlinks=true, allcolors=blue, backref=page]{hyperref}
\usepackage{xspace}
\usepackage{xcolor}
\usepackage{listings}
\usepackage{subcaption}
\newcommand{\degsq}{deg$^2$\xspace}
\newcommand{\faro}{\texttt{faro}\xspace}

\newcommand{\ro}{Rubin Observatory\xspace}
\newcommand{\lsst}{Legacy Survey of Space and Time (LSST)\xspace}
\newcommand{\squash}{SQuaSH\xspace}

\newcommand\arcmin{\mbox{$^\prime$}}%

\definecolor{codegreen}{rgb}{0,0.6,0}
\definecolor{codegray}{rgb}{0.5,0.5,0.5}
\definecolor{codepurple}{rgb}{0.58,0,0.82}
\definecolor{backcolour}{rgb}{0.95,0.95,0.92}

\lstdefinestyle{farostyle}{
    backgroundcolor=\color{backcolour},   
    commentstyle=\color{codegreen},
    keywordstyle=\color{magenta},
    numberstyle=\tiny\color{codegray},
    stringstyle=\color{codepurple},
    basicstyle=\ttfamily\footnotesize,
    breakatwhitespace=false,         
    breaklines=true,                 
    captionpos=b,                    
    keepspaces=true,                 
    numbers=none,                    
    numbersep=5pt,                  
    showspaces=false,                
    showstringspaces=false,
    showtabs=false,                  
    tabsize=2
}
\newcommand\YAMLcolonstyle{\color{black}\mdseries}
\newcommand\YAMLkeystyle{\color{magenta}\mdseries}
\newcommand\YAMLvaluestyle{\color{codepurple}\mdseries}

\lstdefinelanguage{YAML}{
    keywords={true,false,null,y,n},
    keywordstyle=\color{darkgray}\bfseries,
    basicstyle=\YAMLkeystyle,                                 
    sensitive=false,
    comment=[l]{\#},
    morecomment=[s]{/*}{*/},
    commentstyle=\color{black}\ttfamily,
    stringstyle=\YAMLvaluestyle\ttfamily,
    moredelim=[l][\color{black}]{\&},
    moredelim=[l][\color{magenta}]{*},
    moredelim=**[il][\YAMLcolonstyle{:}\YAMLvaluestyle]{:},   
    morestring=[b]',
    morestring=[b]",
    literate =    {---}{{\ProcessThreeDashes}}3
    {>}{{\textcolor{black}\textgreater}}1
    {|}{{\textcolor{black}\textbar}}1
    {\ -\ }{{\mdseries\ -\ }}3,
}

\title{Faro: A framework for measuring the scientific performance of petascale Rubin Observatory data products}

\author[1]{Leanne~P.~Guy}
\affil[1]{Rubin Observatory Project Office, 950 N.\ Cherry Ave., Tucson, AZ  85719, USA}

\author[1,2]{Keith~Bechtol}
\affil[2]{Department of Physics, University of Wisconsin-Madison, Madison, WI 53706, USA}

\author[1]{Jeffrey~L.~Carlin}

\author[1]{Erik~Dennihy}

\author[2]{Peter~S.~Ferguson}

\author[1]{K.~Simon~Krughoff}

\author[3]{Robert~H.~Lupton}
\affil[3]{Department of Astrophysical Sciences, Princeton University, Princeton, NJ 08544, USA}

\author[4]{Colin~T.~Slater}
\affil[4]{University of Washington, Dept.\ of Astronomy, Box 351580, Seattle, WA 98195, USA}

\author[4]{Krzysztof~Findeisen}

\author[3]{Arun~Kannawadi}

\author[3]{Lee~S.~Kelvin}

\author[3]{Nate~B.~Lust}

\author[3]{Lauren~A.~MacArthur}

\author[2]{Michael~N.~Martinez}

\author[3]{Sophie~L.~Reed}

\author[3]{Dan~S.~Taranu}

\author[5]{W.~Michael~Wood-Vasey}
\affil[5]{Department of Physics and Astronomy, University of Pittsburgh, 3941 O'Hara Street, Pittsburgh, PA 15260, USA}

\date{29 June 2022}
 
\begin{document}
\maketitle

\begin{abstract}

The Vera C.\ Rubin Observatory will advance many areas of astronomy over the next decade with its unique wide-fast-deep multi-color imaging survey, the Legacy Survey of Space and Time (LSST)\cite{2019ApJ...873..111I}.
The LSST will produce approximately 20TB of raw data per night, which will be automatically processed by the LSST Science Pipelines to generate science-ready data products -- processed images, catalogs and alerts. 
To ensure that these data products enable transformative science with LSST, stringent requirements have been placed on their quality and scientific fidelity, for example on image quality and depth, astrometric and photometric performance, and object recovery completeness. 
In this paper we introduce \faro, a framework for automatically and efficiently computing scientific performance metrics on the LSST data products for units of data of varying granularity, ranging from single-detector to full-survey summary statistics. 
By measuring and monitoring metrics, we are able to evaluate trends in algorithmic performance and conduct regression testing during development, compare the performance of one algorithm against another, and verify that the LSST data products will meet performance requirements by comparing to specifications. 
We present initial results using \faro to characterize the performance of the data products produced on simulated and precursor data sets, and 
 discuss plans to use  \faro to verify the performance of the LSST commissioning data products.
 
\end{abstract}


\section{Introduction} \label{sec:intro}

Rubin Observatory's \lsst, scheduled to begin operations in 2024, will be the most ambitious and comprehensive optical astronomy survey ever undertaken\cite{2019ApJ...873..111I}.
The LSST will image the entire visible sky twice each week for 10 years, resulting in a dataset that will comprise over 30 trillion observations of 40 billion astronomical sources.
The science enabled by Rubin's LSST will be very broad, ranging from studies of small moving bodies in the solar system to the structure and evolution of the universe as a whole. 

Achieving the ambitious science goals of LSST requires an unprecedented survey dataset and a comprehensive understanding of the survey performance. 
The LSST Science Requirements Document (SRD)\cite{LPM-17} defines a set of ``normative'' science performance metrics that describe the quality of the LSST data products needed to meet these goals, for example on image quality and depth, astrometric and photometric performance, and object recovery completeness.
By defining, measuring, and tracking science performance metrics, we are able to rapidly assess performance, identify changes, problems, or regressions, and take timely action to ensure the survey stays on track to achieve its goals.

As \ro moves into the final stages of construction and begins to prepare for operations, significant effort is being placed on understanding the performance of the as-built system and the quality of the science-ready data products it will deliver.
\faro\footnote{\url{https://github.com/lsst/faro}} is an open-source Python framework distributed as part of the LSST science pipelines\cite{2019ASPC..523..521B,2018PASJ...70S...5B} that computes scalar performance metrics on the outputs of the LSST science pipelines.
It provides a customizable and easily-extensible set of tools to:
\begin{itemize}
\item Generate artifacts to verify the normative science performance metrics detailed in the LSST Science Requirements Document,
\item Compute additional non-normative metrics for science validation, 
\item Monitor performance, carry out regression analysis, and characterization of the LSST science pipelines, 
\item Providing a ``first-look'' data quality analysis capability to inform observatory operations.
\end{itemize}

\section{The Faro Framework} \label{sec:faro}

Two main components form the basis of the \faro framework: a collection of \texttt{Task}s that measure specific metric values and a set of base classes that handle data i/o for various types of input data units corresponding to the granularity of metric computation.
\faro builds upon much of the existing infrastructure of the LSST Science Pipelines, in particular, the ``Generation 3'' middleware, including the Data Butler and Task Framework\cite{SPIE-12189-40}, and the LSST verification framework\cite{DMTN-098, SQR-019}.
The LSST Data Butler (hereafter the ``Butler'') provides an abstracted data access interface that is used to read and write data without knowing the details of file formats or locations.
The Butler organizes data, both raw and processed, into data repositories.
A \texttt{Task} is a reusable unit of code in the LSST science pipelines infrastructure used to process data.
Each \texttt{Task} has a specialized configuration object attached to it and provides a \texttt{run()} method that implements the algorithm to execute. 
Each \faro metric has an associated \texttt{Task} class that implements the mathematical algorithm to compute the scalar metric value on the input data.
The LSST Verification Framework, \texttt{lsst.verify}, is a framework for making verification measurements in the LSST Science Pipelines.
\faro takes as input catalog data products generated by the LSST Science Pipelines\footnote{The LSST science pipelines produce catalog data products both in FITS and parquet file format. 
\faro works exclusively with parquet formatted data.} stored in a Butler repository,
 and computes scalar performance metrics on them. 
The resulting metric values are persisted in the same Butler repository alongside the input data products, each with an associated unique and searchable data identifier,
as \texttt{lsst.verify.Measurement} objects.
\faro has adopted a modular and extensible design that can be configured to run any subset of metrics on any subset of data products, e.g., a subset of the real-time metrics computed on a single exposure.

\subsection{Analysis Contexts} \label{ssec:analysis_context}

An analysis context defines the input data unit corresponding to the granularity of metric computation, e.g, per-detector, per-visit, per-patch, or per-tract.
A tract is a portion of sky, a spherical convex polygon,  within the LSST all-sky tessellation, also known as a sky map. 
Each tract is subdivided into sky patches,  quadrilateral sub-regions of a sky tract, with a size in pixels chosen to fit easily into memory on desktop computers. 
A visit is a single observation of an LSST field and a detector is subdivision of a visit corresponding to single CCD on the LSST focal plane. 
The following analysis contexts are currently supported in \faro:
\begin{itemize}
\item per-visit or per-detector source catalogs, i.e., single-visit detections,
\item per-tract or per-patch object catalogs, i.e., coadd detections, both single-band and multi-band input,
\item per-tract or per-patch matched source catalogs, i.e., set of single-visit detections of the same objects, both single-bandband and multi-band input,
\end{itemize}

\faro supports metric calculation across multiple analysis contexts; the same mathematical function can be called in a different analysis context to produce a different metric value.
For example, the photometric and astrometric repeatability metrics and band-to-band astrometric transformation accuracy metrics are computed at the visit level on single-frame data products, whereas residual PSF ellipticity correlation metrics could be computed at either a per-visit level or on a per-tract level on the coadded data products.
For some metrics, only certain analysis contexts make sense, and often the choice of analysis context will be dictated by statistics, e.g., a small 1\,\degsq  dataset may not contain enough data to compute some metrics on a per-tract level.
\faro allows users to easily define new analysis contexts for their particular science needs.

\subsection{Base Classes} \label{ssec:base_classes}

\faro provides a set of base classes corresponding to the various analysis contexts described above that use the \texttt{Task} framework to build a workflow and interact with the Butler for data i/o.
The base classes abstract away the data i/o, leaving the scientist free to focus on the algorithmic details  relevant to their particular analysis.
The primary base classes in the lsst.faro package are \texttt{CatalogMeasurementBaseConnections}, to define the desired i/o, \texttt{CatalogMeasurementBaseConfig}, to provide the configuration, and \texttt{CatalogMeasurementBaseTask}, to run the algorithm and store the output data in the Butler.
Each of these base classes respectively inherits from \texttt{MetricConnections}, \texttt{MetricConfig}, and \texttt{MetricTask}, in the \texttt{lsst.verify} package and adds additional functionality for computing science performance metrics based on Source and Object catalog inputs.
Additional subclasses can be easily added following this scheme for new analysis contexts.

\subsection{Stages of Metric Computation} \label{ssec:stages}

Metric computation with \faro proceeds through three stages:
\begin{enumerate}
\item \textbf{Preparation:} any intermediate data products that are needed as input to the subsequent measurement step are assembled and persisted in the Butler.
\item \textbf{Measurement:} the metric \texttt{Task} is run to compute a measurement for each unit of data (i.e., a quantum of processing with a particular a dataId for the output measurement). Measurements are stored as \texttt{lsst.verify.Measurement} objects and persisted in the Butler.
\item \textbf{Summary:} a single scalar summary statistic, e.g., a mean or median,  is generated from the collection of input measurements computed in the measurement stage. Summary statistics are stored as \texttt{lsst.verify.Measurement} objects and persisted in the Butler.
\end{enumerate}
Not all metrics will necessarily have a preparation stage as there may be no additional intermediate data products needed to compute the metric than those generated as part of the pipelines processing. However, all must have a measurement and summary stage.

For example, the photometric repeatability metric, PA1, measures the RMS photometric repeatability of bright non-saturated point sources in a single filter.
During the preparation stage, a matched source catalog is created for each tract and band that matches source detections in individual visits that correspond to the same physical astronomical object\footnote{LSST defines an Object as an astrophysical physical object, e.g., a star, galaxy or asteroid, and a Source as a single detection of an astrophysical object in an image. The association of non-moving Sources leads to Objects; the association of moving Sources leads to Solar System Objects.}.
During the measurement stage, for each tract and band, the matched catalog computed and stored in the preparation stage is loaded into memory and used to compute the RMS scatter of fluxes for a single astrophysical object. 
In the final summary stage, the measurements for the ensemble of individual tracts computed in the measurement stage are loaded to compute a median summary statistic per band. 
This final summary statistic characterizes the overall performance for the dataset per band and is stored as an \texttt{lsst.verify} object in the Butler. 

\subsection{Processing Pipelines} \label{ssec:pipelines}

A \texttt{Pipeline} is a collection of \texttt{PipelineTask}s, specified in the form of YAML files that that can be run in parallel or serial.
A \texttt{PipelineTask} is a special case of a \texttt{Task} that can read inputs from and write outputs to a Butler.
Pipelines are used to define the \faro metrics to be computed together with the detailed execution parameters for metric calculations.
The \faro package provides a set of pre-defined pipelines, defined per analysis context, for many common configurations of computing metrics. 
\faro pipelines can be run as an afterburner to the LSST Science Pipelines or can be interspersed with science pipelines processing to compute, as long as the necessary input data products have been computed and are stored in the Butler.
For example, all single-visit metrics can already be computed once the single-frame processing steps have completed. 
This enables us to promptly assess performance and take any corrective action early. 
Fig.~\ref{fig:faro_pipeline} shows an example of a \faro pipeline describing the measurement stage of the computation of two performance metrics using a per-tract matched catalog analysis context.
Pipelines can be built hierarchically, with a high-level pipeline calling lower-level pipelines.

\begin{figure}[ht]
  \begin{center}
  \lstset{language=YAML}
  \begin{lstlisting}
description: Compute metrics from matched catalogs
tasks:
  PA1:
    class: lsst.faro.measurement.TractMatchedMeasurementTask
    config:
      connections.package: validate_drp
      connections.metric: PA1
      python: |
        from lsst.faro.measurement import PA1Task
        config.measure.retarget(PA1Task)
  PF1_design:
    class: lsst.faro.measurement.TractMatchedMeasurementTask
    config:
      connections.package: validate_drp
      connections.metric: PF1_design_gri
      python: |
        from lsst.faro.measurement import PF1Task
        config.measure.retarget(PF1Task)
        config.measure.threshPA2 = 15.0
    \end{lstlisting}
    \end{center}
  \caption{\label{fig:faro_pipeline} 
  A \faro pipeline describing the measurement stage of the computation of two performance metrics per-tract from matched catalogs.}
\end{figure}

\section{Adding a  Metric} \label{sec:add}

We anticipate that many developers and scientists will want to define their own metrics specifically to address concerns that arise as LSST progresses through the construction phase and into operations.
The first step in writing a new metric is to choose the analysis context (\S~\ref{ssec:analysis_context}) and to review the associated \texttt{Connections}, \texttt{Config}, and \texttt{Task} base classes (\S~\ref{ssec:base_classes}) to understand the in-memory Python objects that will be passed to the run method of the metric measurement task, and the configuration options.
If the required analysis context and associated base class do not exist, they must first be created. 
Once the analysis context and base class are defined, the metric can be implemented by creating a measurement \texttt{Task}, an instance of \texttt{lsst.pipe.base.Task} that will operate on in-memory Python objects and compute the metric. 
Finally, the new metric is added to the appropriate pipeline YAML file so that it can be run.
Fig.~\ref{fig:num_sources_task} shows an example of a non-normative metric \texttt{Task} written to compute the number of entries in an input Source or Object catalog, \texttt{NumSourcesTask}

\begin{figure}[ht]
  \begin{center}
  \lstset{language=python}
  \begin{lstlisting}
  
class NumSourcesTask(Task):

    ConfigClass = NumSourcesConfig
    _DefaultName = "numSourcesTask"

    def run(self, metricName, catalog, **kwargs):
        self.log.info("Measuring %s", metricName)
        if self.config.doPrimary:
            nSources = np.sum(catalog[self.config._getColumnName(
                "detect_isPrimary")])
        else:
            nSources = len(catalog)
        self.log.info("Number of sources (nSources) = %i", nSources)
        meas = Measurement(metricName, nSources * u.count)
        return Struct(measurement=meas)
  
    \end{lstlisting}
    \end{center}
  \caption{\label{fig:num_sources_task}
  Implementation of a simple task to count the number of Sources or Objects in an input catalog. }
  \par\medskip
\end{figure}

\section{Tracking and Visualization} \label{sec:tracking}

\begin{figure}[ht]
  \par\medskip 
  \centering
  \includegraphics[width=0.98\textwidth]{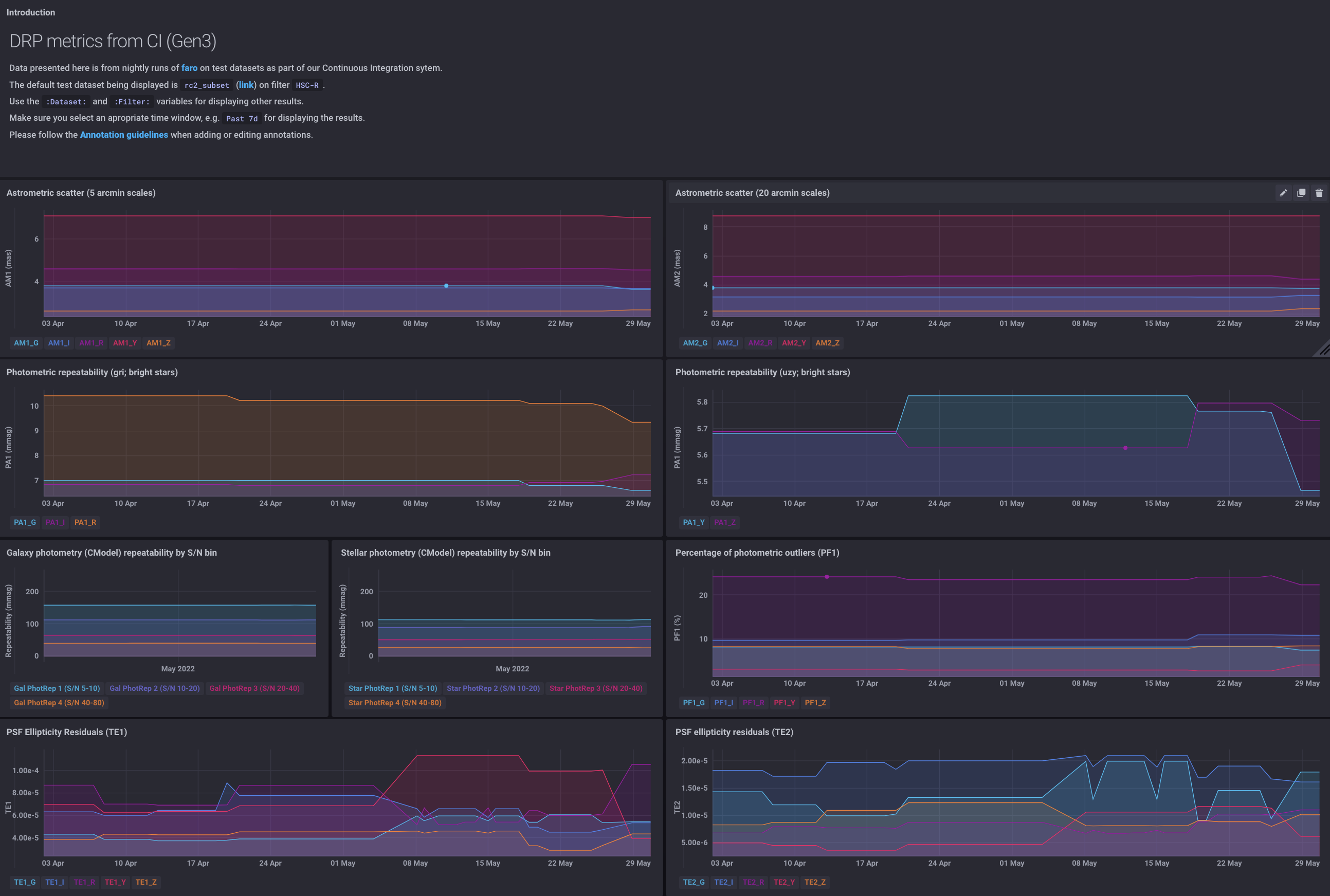}
  \par\medskip   
  \caption{\label{fig:squash_metrics_ci} 
  InfluxDB/Chronograph dashboard showing the time evolution of a sample of metrics computed as part of the nightly builds on a small subset of the HSC RC2 dataset.}
\end{figure}

We visualize and track the time evolution of \faro metrics using the LSST metrics dashboard, \squash\cite{SQR-009}.
\squash makes use of the InfluxDB\footnote{\url{https://www.influxdata.com/}} time series database as the backend store for time-stamped metrics and the Chronograf\footnote{\url{https://www.influxdata.com/time-series-platform/chronograf/}} user interface.
InfluxDB is designed to store large volumes of time series data and quickly perform real-time analysis on that data.
Chronograf provides dashboards to easily visualize and track metrics stored in InfluxDB.
Metrics computed by \faro on the outputs of the LSST Science Pipelines are stored in the Butler and then pushed to the \squash using the InfluxDB REST API.
From the Chronograf UI, the user can query the metric values, aggregate results, and create alerts to measured metrics that change or go out of specification.
By tracking metric measurements, we are able to understand trends in the algorithmic performance of the LSST Science Pipelines and ultimately verify that we will meet our requirements.
We have configured several dashboards for tracking \faro metrics computed on various precursor datasets at different cadences, e.g., nightly, monthly.
Fig.~\ref{fig:squash_metrics_ci} shows the dashboard for \faro metrics computed on a small subset of the HSC RC2 dataset (\S \ref{sssec:ci}). 

\section{Faro Applications} \label{sec:applications}

\subsection{Monitoring Science Pipelines Performance} \label{ssec:monitoring}

\faro is used to monitor the scientific performance of the LSST science pipelines on precursor datasets of different sizes and regularly processed on a variety of cadences\cite{dmtn-091}.
By tracking metrics computed on the outputs of evolving versions of the science pipelines using a fixed dataset, we can rapidly identify the effect of code or configuration changes on the data products as development progresses during the LSST construction period.

\subsubsection{Nightly Integration}\label{sssec:ci}

\begin{figure}[ht]
\begin{subfigure}{.5\textwidth}
    \centering
    \includegraphics[width=0.98\textwidth]{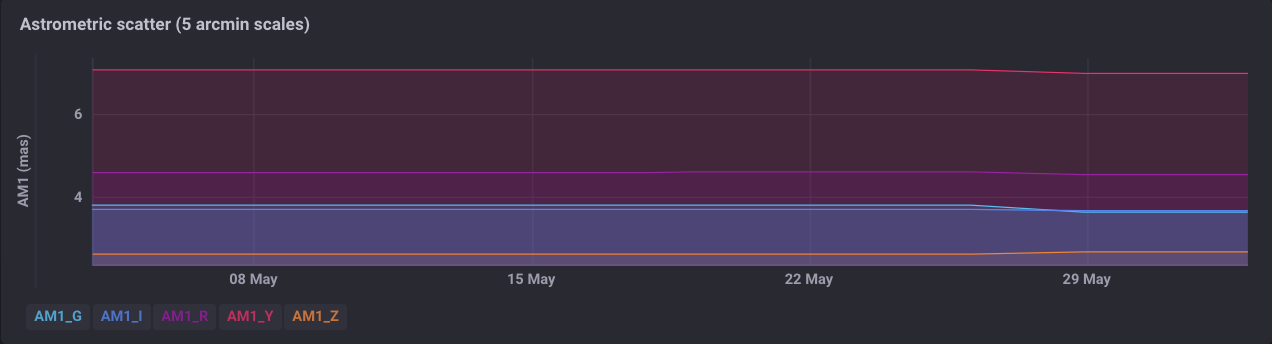}
\end{subfigure}
\begin{subfigure}{.5\textwidth}
    \centering
    \includegraphics[width=0.98\textwidth]{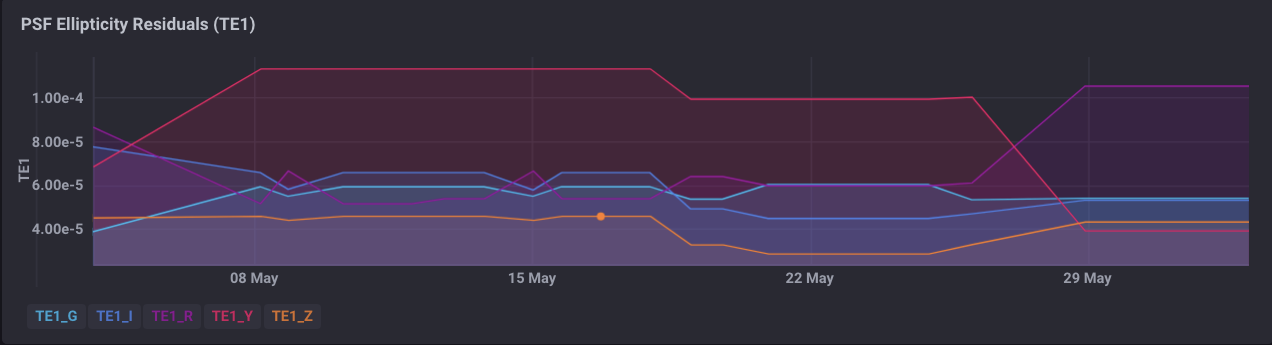}
\end{subfigure}
\par\medskip 
\begin{subfigure}{.5\textwidth}
    \centering
    \includegraphics[width=0.98\textwidth]{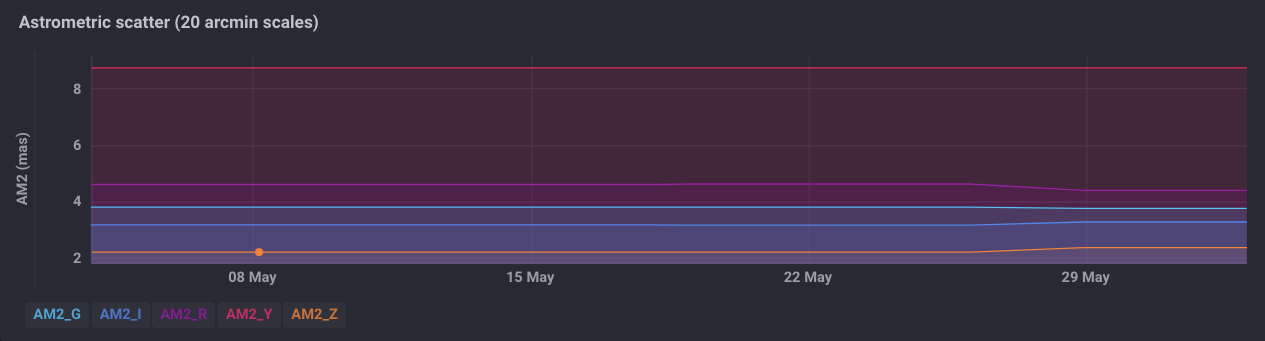}
     \caption[\small]{(a) Astrometric precision metrics}
\end{subfigure}
\begin{subfigure}{.5\textwidth}
    \centering
    \includegraphics[width=0.98\textwidth]{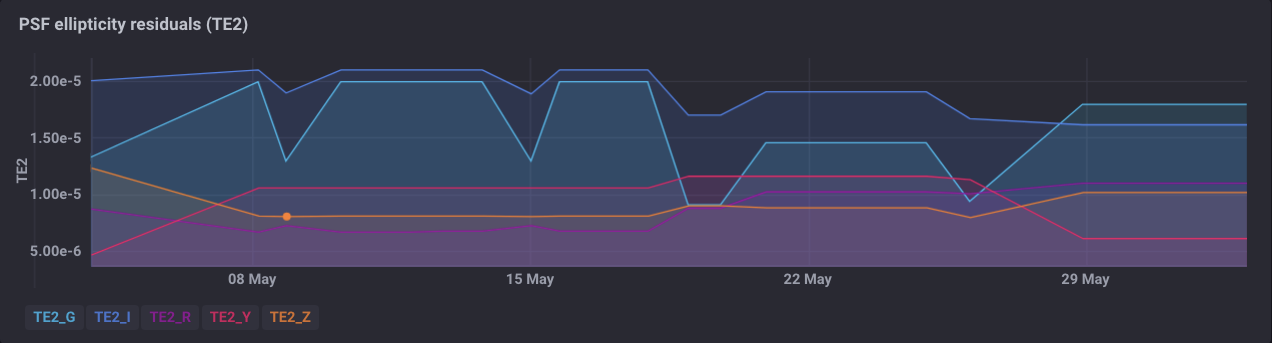}
    \caption[\small]{(b) Residual PSF ellipticity correlation metrics}
\end{subfigure}
\par\medskip 
\caption{\label{fig:ci_metrics}
Examples of the time evolution of four \faro metrics computed on a nightly cadence over a period of one month on the outputs of the nightly data release processing of a small subset of the HSC PDR2 dataset.  
Left shows the AM1 metric that characterizes the astrometric scatter on two different spatial scales, 5 and 20 arcmins, and right shows the residual PSF ellipticity correlations, TE1 and TE2. }
\end{figure}

The LSST Science pipelines are run nightly and weekly on a small subset of the Hyper Suprime-Cam (HSC) Release Candidate 2 (HSC RC2) dataset\footnote{\url{https://github.com/lsst-dm/rc2_subset}} in the LSST Jenkins-based continuous integration (CI) system\cite{2018SPIE10707E..09J}.
This dataset consists of the central 6 detectors for 8 randomly chosen visits in the 5 broad band filters in the HSC COSMOS field and was produced specifically for measuring metrics in a CI environment.
\faro pipelines are run on the outputs in a per-visit or per-detector analysis context.
Given the small size of the dataset, there are not enough statistics to compute metrics on a per-tract level.
The goal is to track daily the performance of the pipelines and the quality of the data products and to rapidly identify the effect of code or configuration changes on the data products as development progresses during the LSST construction period.
Fig.~\ref{fig:ci_metrics} shows an example of two tracked metrics and how they were used to identify the effect of a software change.
Metrics are reviewed regularly allowing us to identify changes and understand the impact of changes in the science pipelines.

\subsubsection{Software Release Characterization } \label{sssec:characterization}

\begin{figure}[ht]
  \centering
  \includegraphics[width=0.45\textwidth]{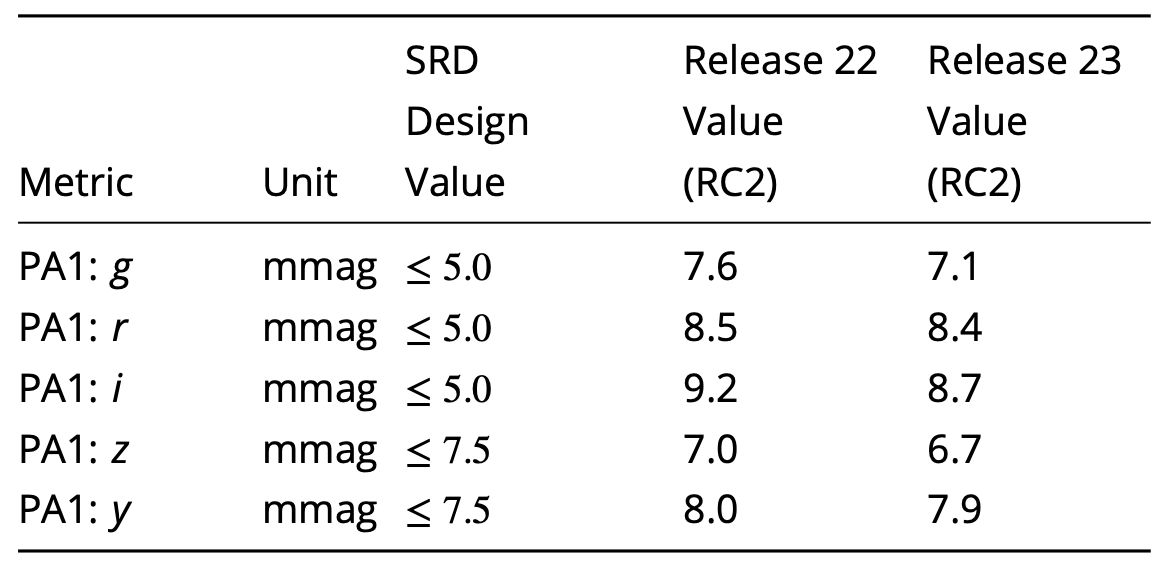} 
  \hspace{0.5cm}
  \includegraphics[width=0.45\textwidth]{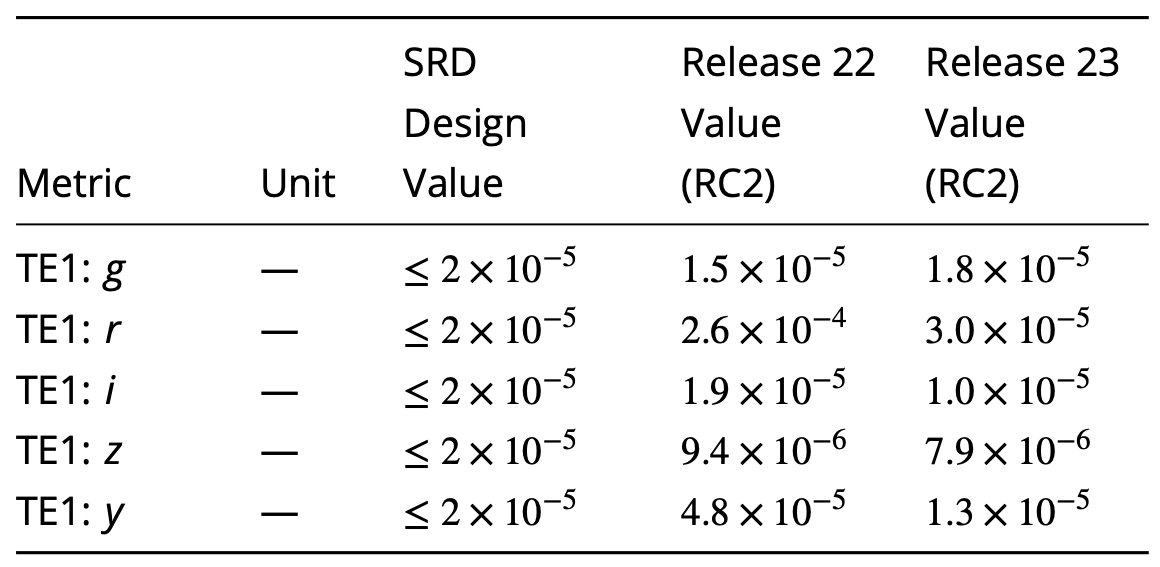}
  \par\medskip
  \caption{\label{fig:cmr_r23}
  Excerpt from the  LSST Science Pipelines major Release 23 Characterization Metric Report on the HSC RC2 dataset. Metric values are compared with those of the previous release of the science pipelines and with the design specifications from the LSST Science Requirements document SRD.}
\end{figure}

Major releases of the LSST science pipelines are made approximately once every six months. 
Every major release is accompanied by a Characterization Metric Report, which describes the scientific performance of the release by computing \faro metrics on the full HSC RC2 dataset. 
HSC RC2 consists of 3 tracts of data selected to test various pathological cases, e.g., difficult astrometric solutions, extremely good seeing yielding an under-sampled PSFs, difficult fields for deblending, and large galaxies, among others.
These three tracts each contain 112--149 visits split between the HSC-G, HSC-R, HSC-I, HSC-Z, and HSC-Y (grizy) filters.
The Characterization Metric Report compares values of metrics computed on the HSC RC2 dataset with a) those computed using the previous release of the science pipelines and b) the LSST SRD design specifications. 
This allows us to monitor the effects of major strategic developments in algorithms over longer periods.
Fig.~\ref{fig:cmr_r23} shows an excerpt from the Characterization Metric Report \cite{dmtr-351} for the most recent release of the LSST science pipelines, Release 23.0.1.
From this excerpt, we see that there were improvements in the photometry and ellipticity correlation metrics as compared to the previous release, which reflect improvements in the photometric calibration algorithms to better handle fields of view near the spatial edges of survey footprints. 

\subsection{Rubin Auxiliary Telescope Imaging Surveys} \label{ssec:auxtel}

\begin{figure}[ht]
\begin{subfigure}{.5\textwidth}
    \centering
    \includegraphics[width=1.0\textwidth]{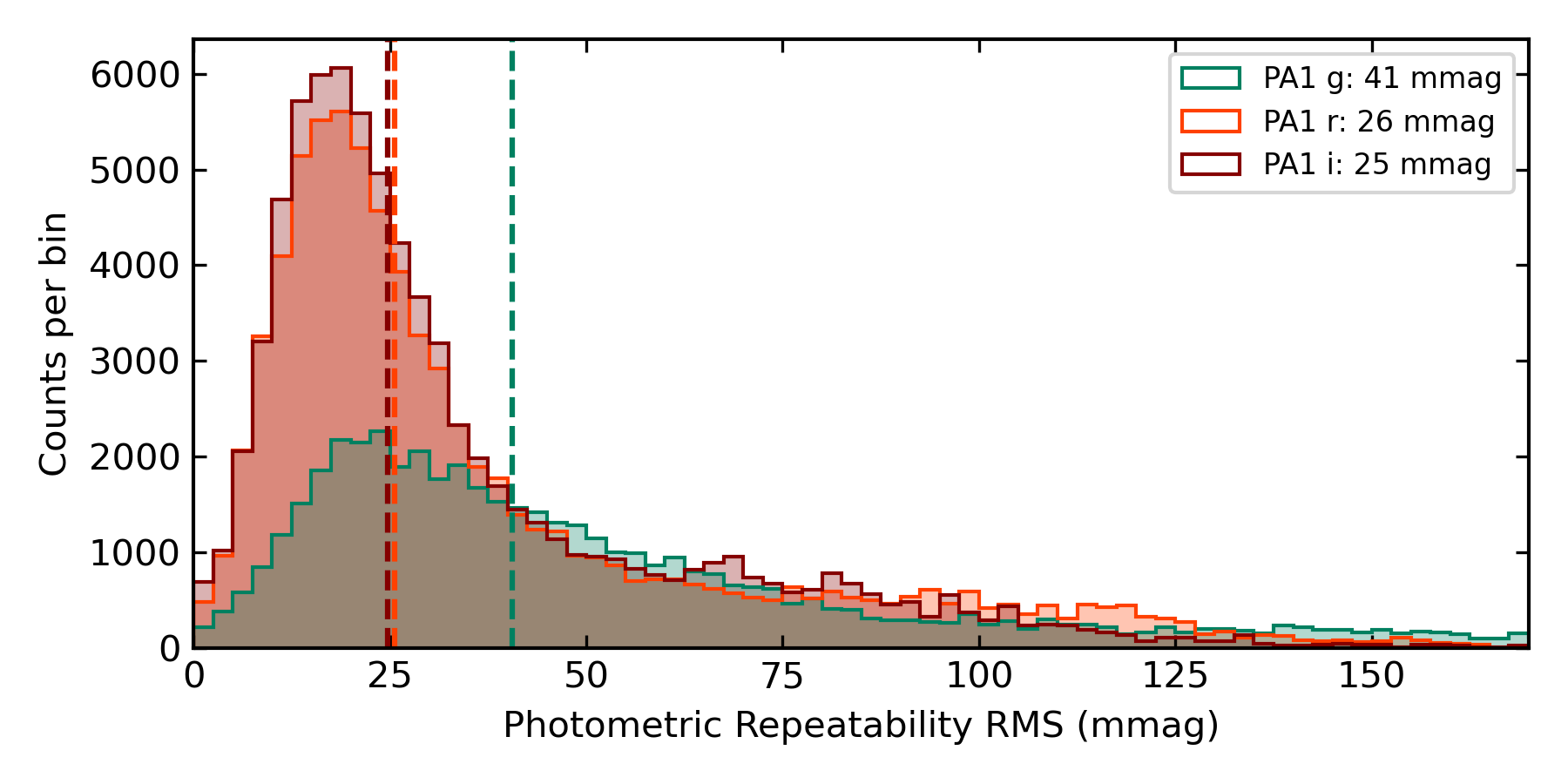}
     \caption{(a) full detector}
\end{subfigure}
\begin{subfigure}{.5\textwidth} 
    \centering
    \includegraphics[width=1.0\textwidth]{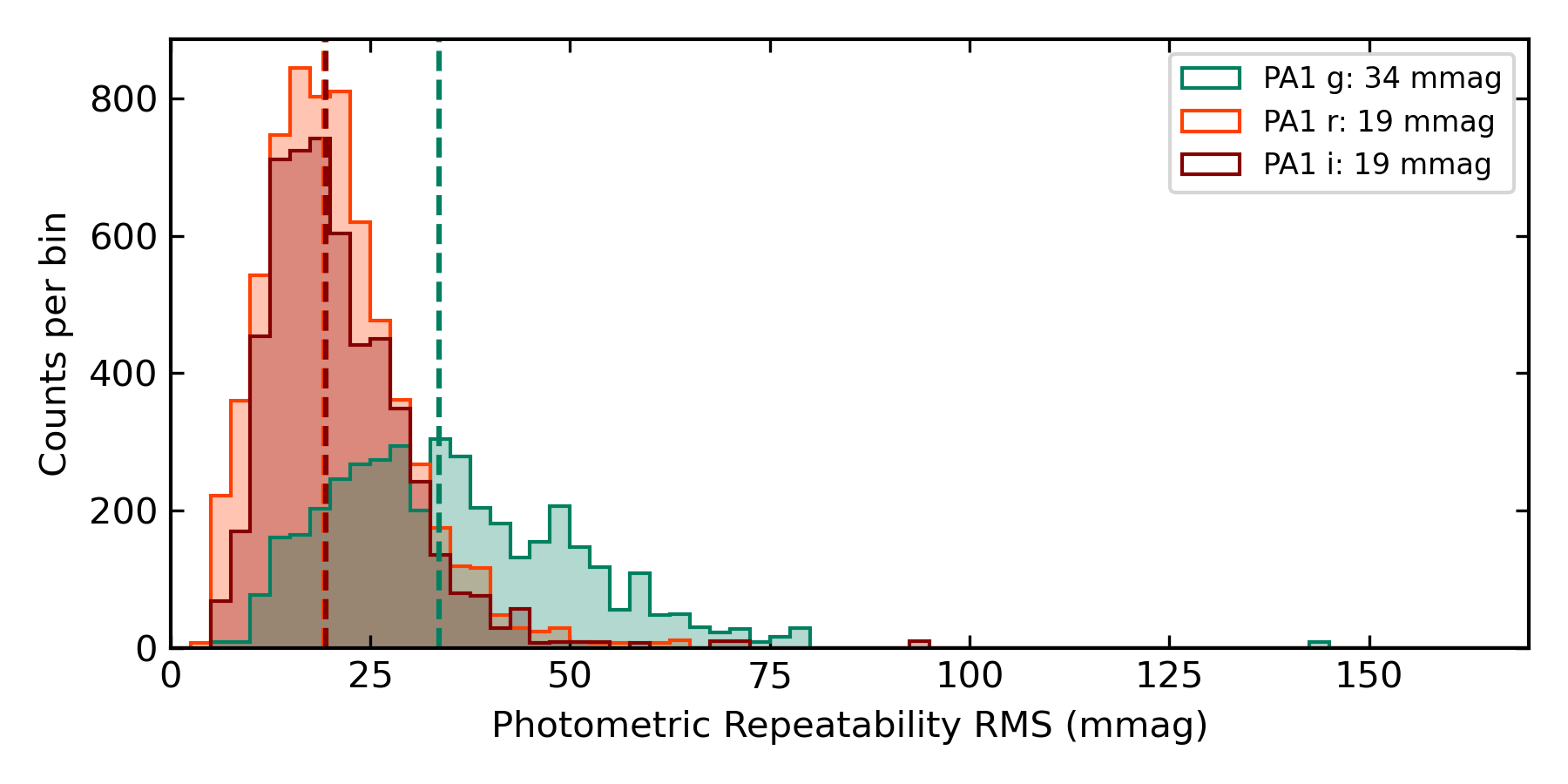}
    \caption{(b) central detector region}
\end{subfigure}
\par\medskip 
\caption[short]{\label{fig:faro_auxtel_metrics} 
Distributions of RMS values of the photometric repeatability of  bright non-saturated  point sources detected as part of the initial AuxTel imaging campaign. Vertical dashed lines show the calculated PA1 metric value for each bandpass. The left panel shows the distribution of PA1 values for all sources detected and the right panel includes only sources in the central region of the detector (1000 < X,Y < 3000 pixels). The central region of the detector is not affected by the vignetting due the epoxy creep on the filters leading to much tighter distributions and lower PA1 values.}
\end{figure}

The Rubin Auxiliary Telescope\cite{10.1117/12.2561112} (AuxTel) is a 1.2m f/18 telescope that sits adjacent to Rubin Observatory that will be used to measure atmospheric transmission using broadband spectroscopy of bright stars throughout survey operations. 
During commissioning, AuxTel is also being used as a control systems and data processing pathfinder via a series of imaging surveys. The imaging surveys are being conducted in three bands (\emph{g}, \emph{r}, and \emph{i}) using a 2x30s exposure sequence per visit.
The first AuxTel imaging campaign collected multi-band data over a 1\,\degsq field of high stellar density in a series of 110 pointings, completing an average of 10 visits per pointing between Feb 2022 and May 2022. The data was processed with the LSST Science Pipelines, including measuring science performance metrics with \faro. This dataset is sufficient to explore coadds, repeatability, and optimize processing configurations.
Single-visit and matched-visit science performance metrics, including the photometric repeatability PA1, were computed with \faro using data generated from the AuxTel Imaging Survey.
Fig.~\ref{fig:faro_auxtel_metrics} shows the PA1 metric values for bright non-saturated  point sources per bandpass separated into two distributions (a) sources detected across the full focal plane and (b) only sources detected in the center  region (1000 < X,Y < 3000 pixels). The physical filters used in the initial imaging campaign suffer from the inward creep of leaked epoxy around their edges, resulting in time-evolving vignetting. This effect is readily seen in the PA1 metric calculation.

\subsection{Rapid and first-look analysis} \label{ssec:rapid}

During LSST observing, early analysis of the images coming off the camera will be carried out on the mountain on the timescales of a few minutes. 
The following are examples of metrics that could be implemented in \faro to support this rapid first-look analysis:
\begin{itemize}
\item Sky brightness, counts, and variance
\item PSF FWHM and PSF ellipticity
\item Comparison to reference catalogs, e.g., Gaia to look for an excess/deficit of detections, astrometric offsets, photometric residuals, or problems with PSF modeling,
\item Effective survey depth
\end{itemize}

\subsection{Rubin Data Previews} \label{ssec:datapreviews}

\begin{figure}[h]
\begin{subfigure}{.5\textwidth}
    \centering
    \includegraphics[width=0.98\textwidth]{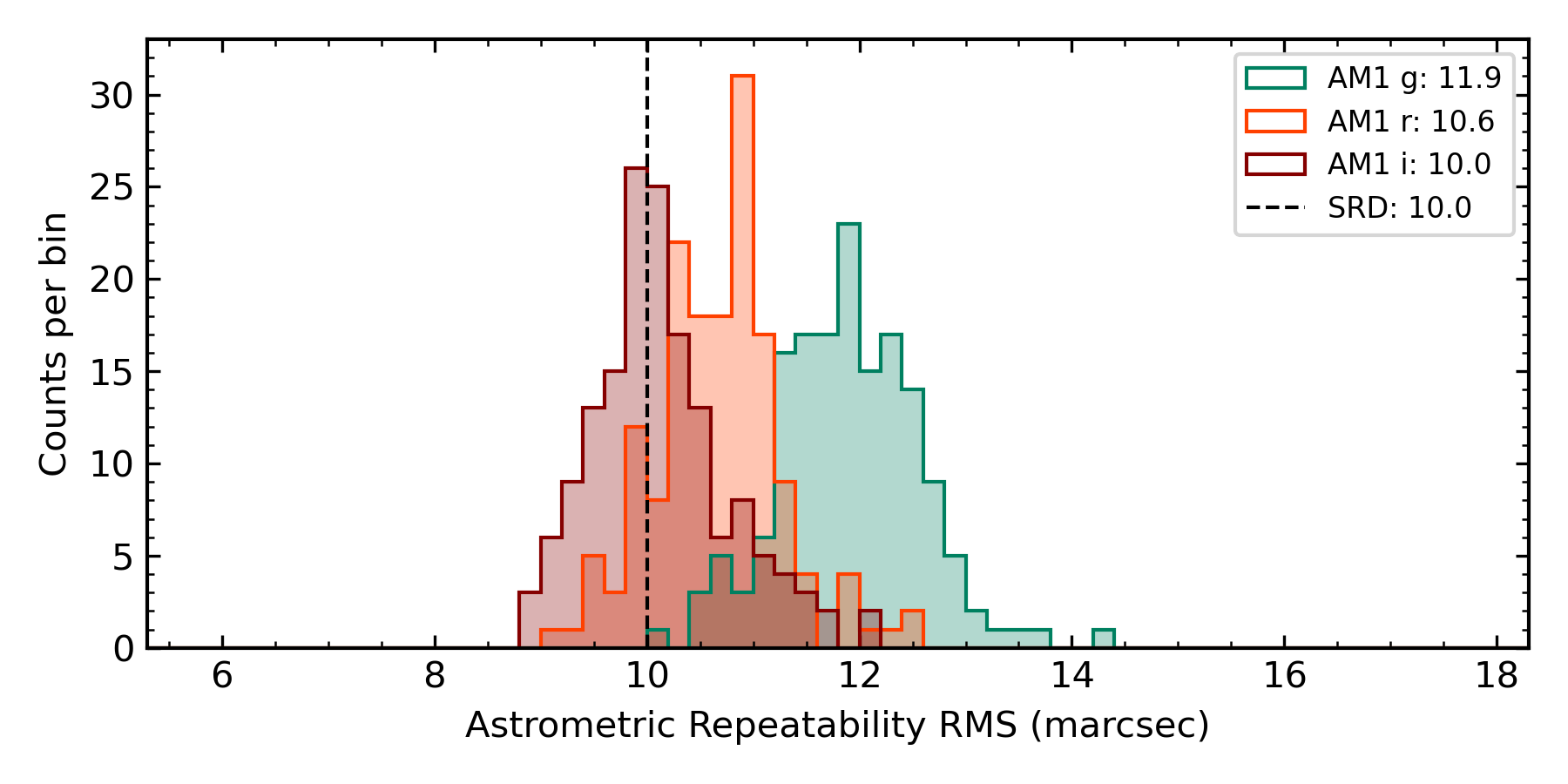}
\end{subfigure}
\begin{subfigure}{.5\textwidth}
    \centering
    \includegraphics[width=0.98\textwidth]{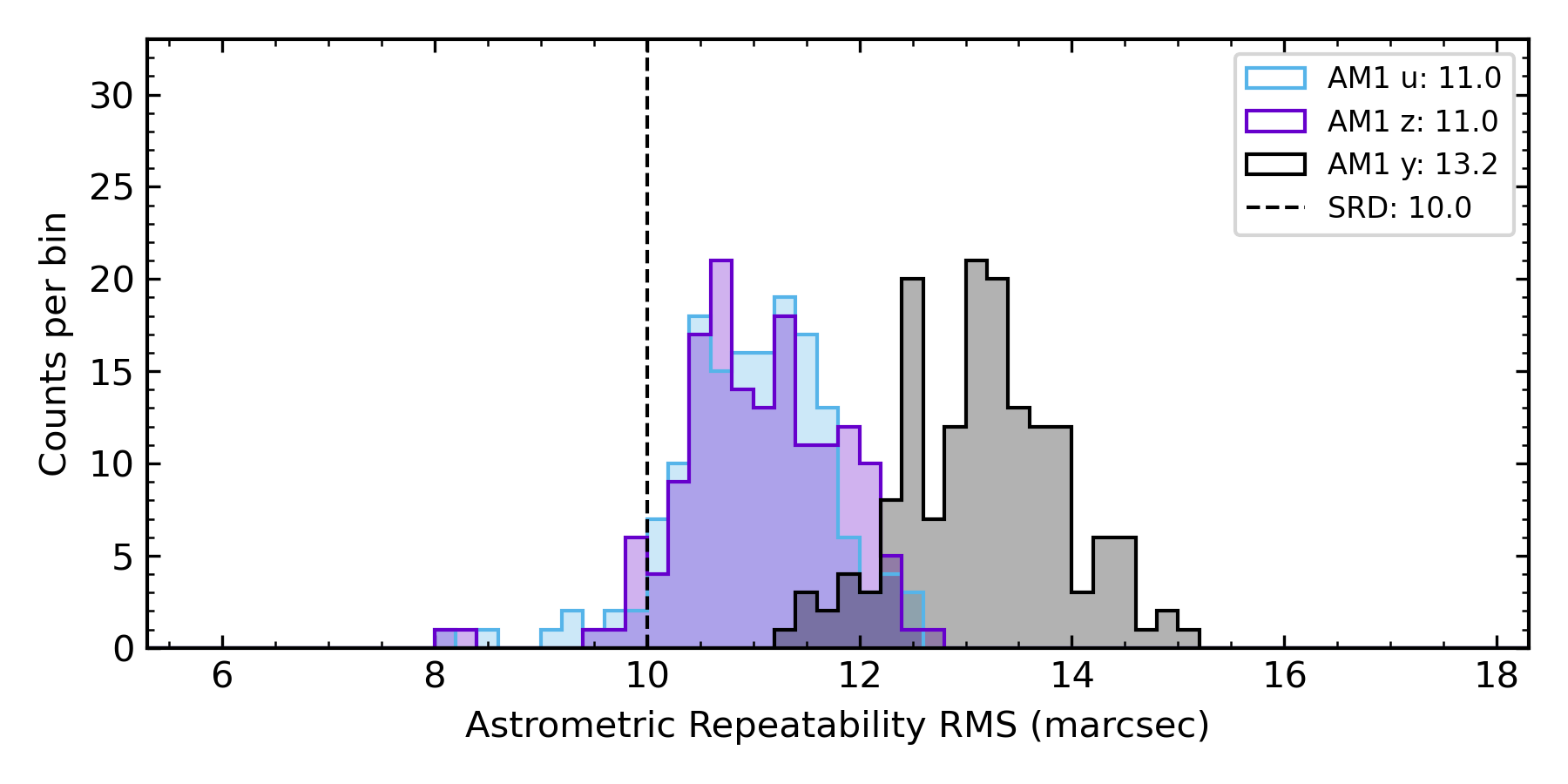}
\end{subfigure}
\par\medskip
\caption{\label{fig:faro_dp02_distr_am1}
Distributions of the RMS values of the astrometric repeatability of bright non-saturated point sources detected as part of the DP0.2 processing campaign on 5\arcmin\, timescales  in the a) \emph{g}, \emph{r}, and \emph{i}  and  b) \emph{u}, \emph{z}, and \emph{y} bands.
The vertical dashed lines show the LSST SRD design specifications. }
\end{figure}

\begin{figure}[h]
\begin{subfigure}{.5\textwidth}
    \centering
    \includegraphics[width=0.98\textwidth]{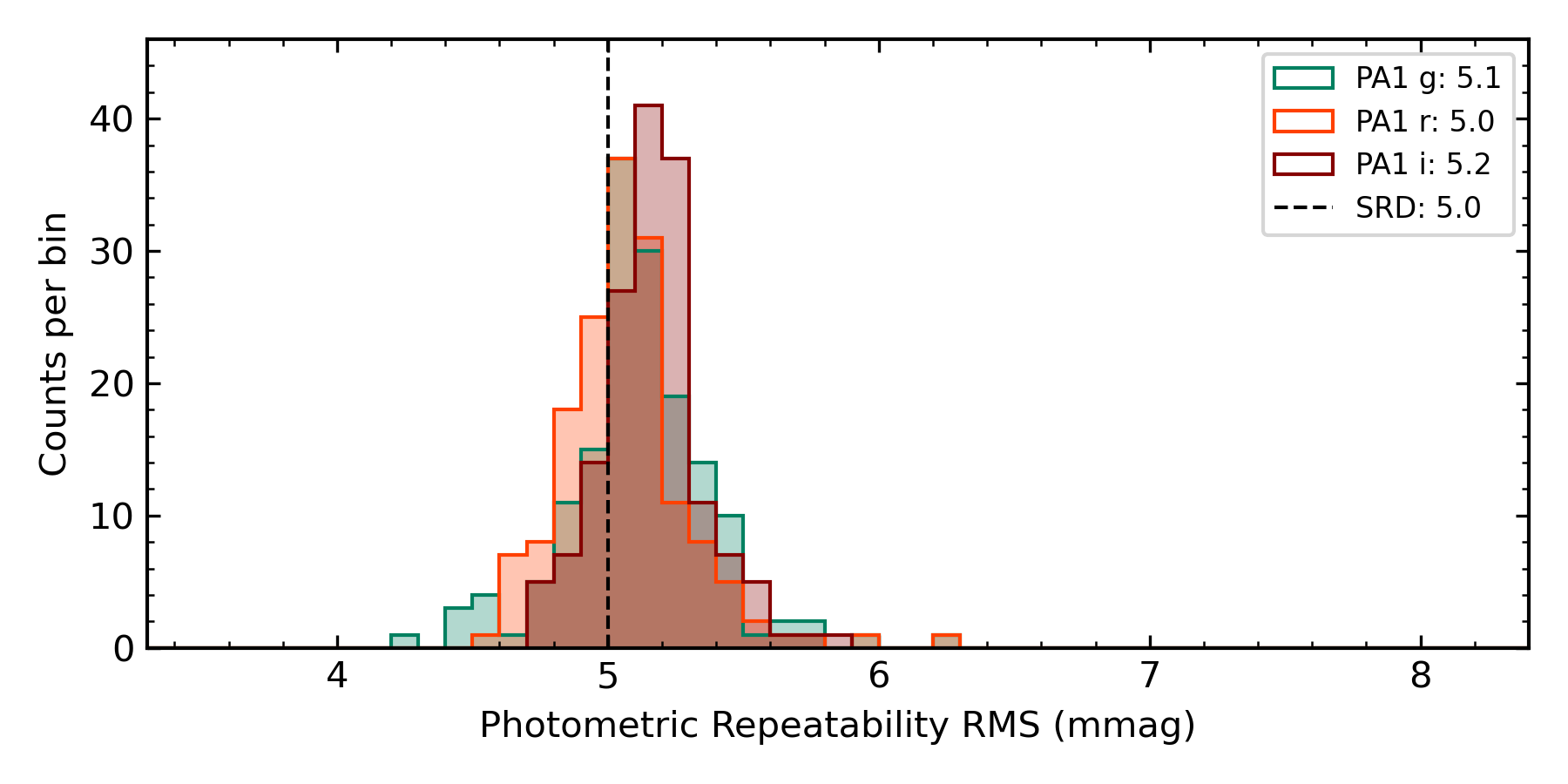}
\end{subfigure}
\begin{subfigure}{.5\textwidth}
    \centering
    \includegraphics[width=0.98\textwidth]{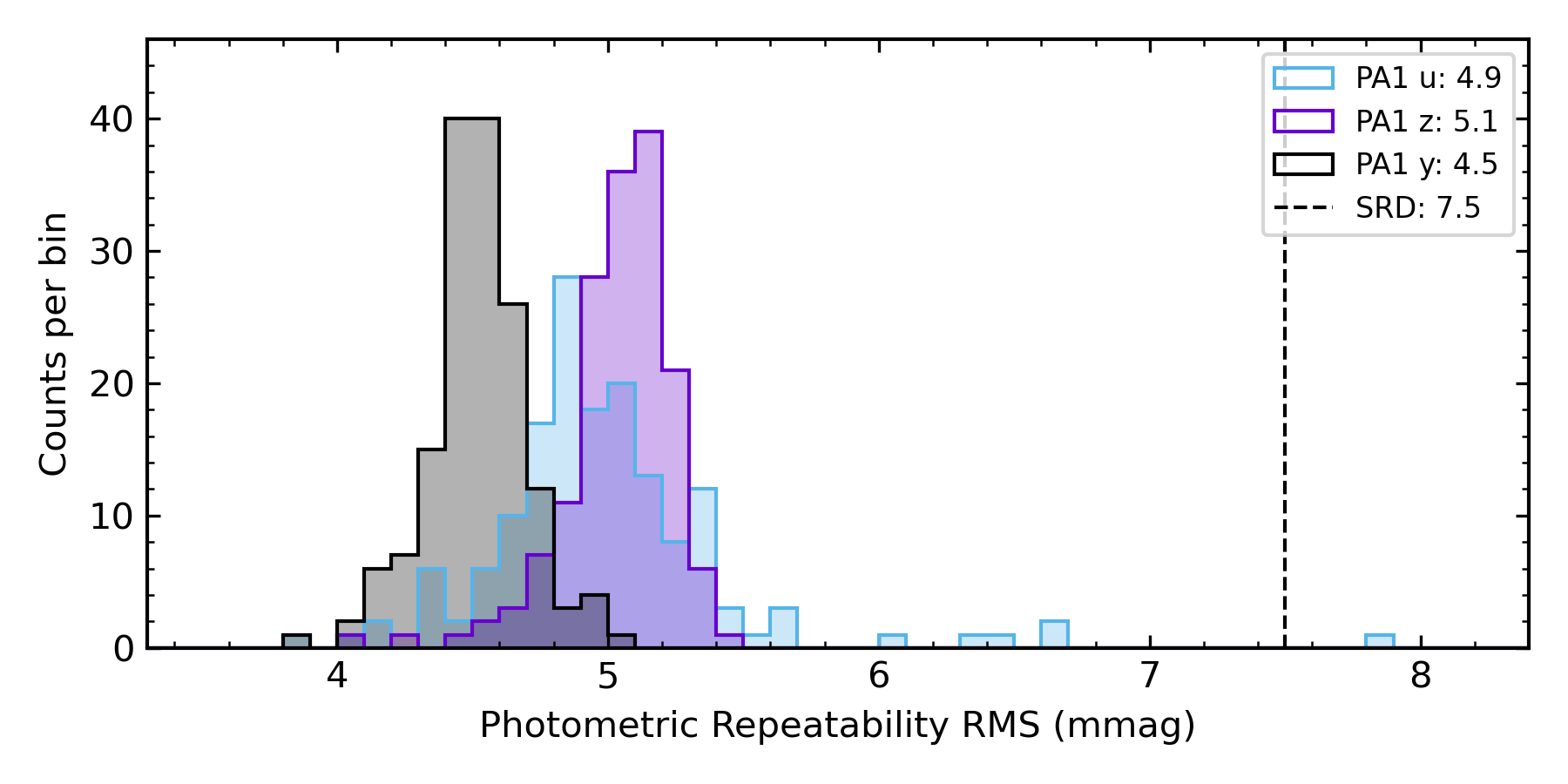}
\end{subfigure}
\par\medskip
\caption{\label{fig:faro_dp02_distr_pa1}
Distributions of the RMS values of the photometric repeatability of bright non-saturated point sources detected as part of the DP0.2 processing campaign in the a) \emph{g}, \emph{r}, and \emph{i}  and  b) \emph{u}, \emph{z}, and \emph{y} bands.  
The vertical dashed lines show the LSST SRD design specifications. }
\end{figure}

\begin{figure}[h!]
\begin{subfigure}{.5\textwidth}
    \centering
    \includegraphics[width=0.98\textwidth]{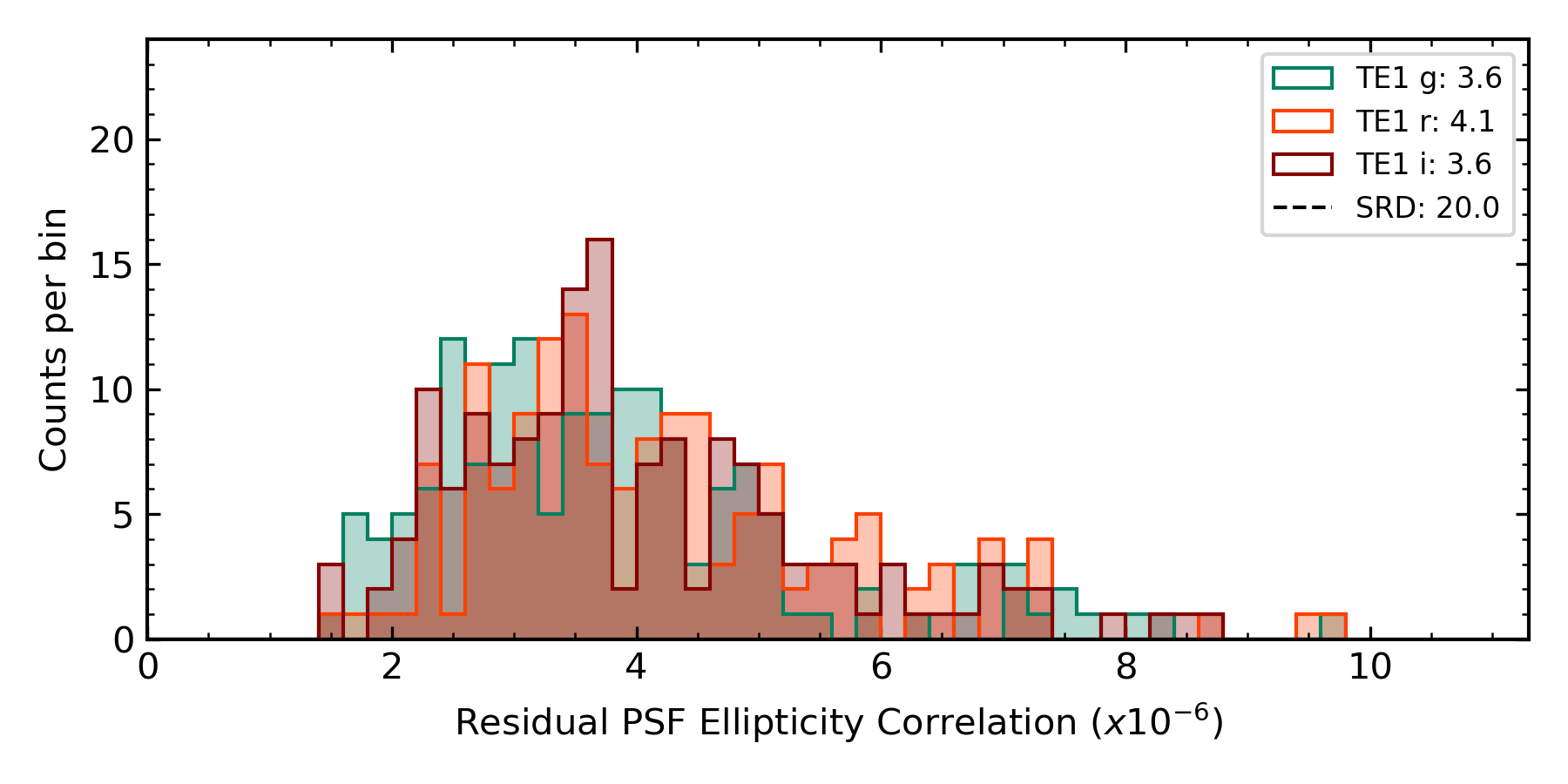}
\end{subfigure}
\begin{subfigure}{.5\textwidth}
    \centering
    \includegraphics[width=0.98\textwidth]{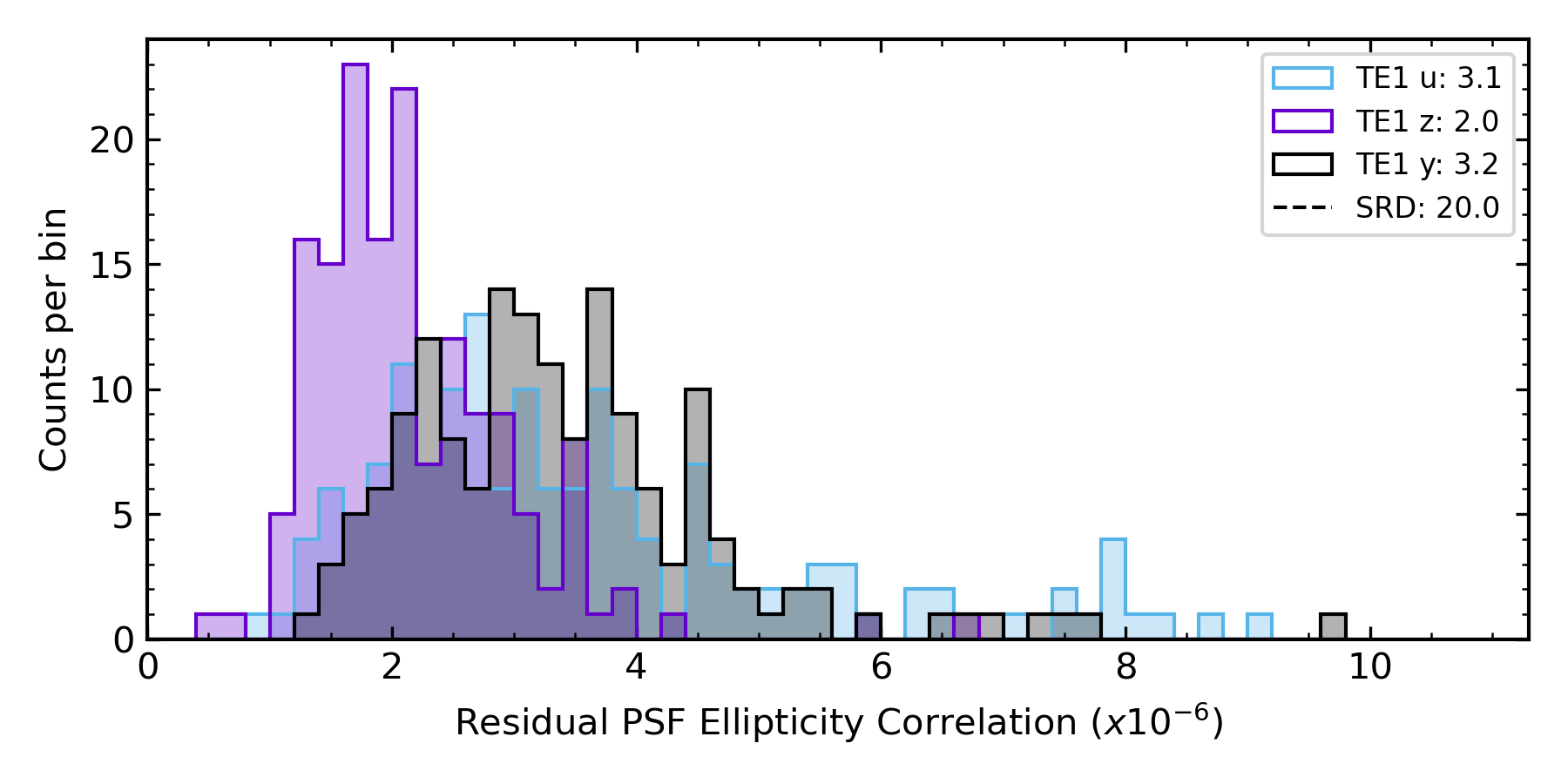}
\end{subfigure}
\par\medskip
\caption{\label{fig:faro_dp02_distr_te1}
Distributions of the residual PSF ellipticity correlations of bright non-saturated  point sources  detected as part of the DP0.2 processing campaign on scales less than or equal to 1\arcmin\, timescales in the a) \emph{g}, \emph{r}, and \emph{i} and  b) \emph{u}, \emph{z}, and \emph{y} bands.
The LSST SRD design specification for this metric is beyond the scales of the plots.}
\end{figure}

During the period leading up to the start of operations in 2024, Rubin Observatory will deliver a series of three Data Previews to the community. 
The goals of these Data Previews are twofold, to serve as an early integration test of the LSST Science Pipelines and the Rubin Science Platform (RSP)\footnote{A set of integrated web applications and services deployed at Rubin Data Access Centers (DACs) through which the scientific community will access, visualize, subset and perform next-to-the-data analysis of LSST Data products.} \cite{lse-319}, and to enable a limited number of scientists to begin early preparations for science with the LSST.
All  Rubin Data Previews are hosted at the Interim Data Facility (IDF) on Google cloud\cite{2021arXiv211115030O}.
Data Preview 0 (DP0) \cite{RTN-001} is the first of these three data previews.
The data set adopted for DP0 is the DESC 300\,\degsq catalog of simulated LSST-like images and catalogs generated by the Dark Energy Science Collaboration (DESC) for their Data Challenge 2 (DC2)\cite{2021ApJS..253...31L}.
As part of the second phase of Data Preview 0, DP0.2, the DESC DC2 simulated images were reprocessed with Release 23 of the LSST Science Pipelines and released to the community on 7 July 2022.
\faro pipelines were run as part of the DP0.2 reprocessing, interspersed with the DRP pipelines to compute performance metrics. 
Fig.~\ref{fig:faro_dp02_distr_am1}, Fig.~\ref{fig:faro_dp02_distr_pa1} and Fig.~\ref{fig:faro_dp02_distr_te1} shows distributions of some of the performance metrics computed as part of DP0.2 together with a comparison to the SRD specifications. 

\section{Current Status and Future Development} \label{sec:future}

The \faro framework is released as part of the LSST Science Pipelines distribution and is running in nightly builds on small representative datasets and on a few tracts of HSC and DESC DC2 data to track the performance of the LSST Science Pipelines.
\faro is used to produce a detailed Characterization Metric Report for major 6-monthly releases of the pipelines. 
\faro is starting to be used to characterize the performance of preliminary data from the Rubin Auxiliary Telescope imaging campaigns and was successfully run at scale on the outputs of the Rubin Operations Data Preview 0.2 processing of the DESC DC2 300\,\degsq dataset.

Many LSST SRD metrics have been implemented in \faro spanning a range of analysis contexts; a near-term focus will be on completing the implementation of the normative science performance metrics.
The next goal will focus on the needs of LSST commissioning and supporting the implementation of ad-hoc metrics to understand the performance of the as-built system, in particular, to provide near real-time feedback and monitoring of the quality of data taken on the mountain.
One indicator of the success of \faro is the steadily increasing number of active developers defining and implementing metrics to characterize and diagnose performance and issues.

\section{Conclusions} \label{sec:conclusions}

In this paper, we have presented \faro, the LSST framework for automatically computing scalar performance metrics on the outputs of the LSST science pipelines on a range of spatial and temporal scales.
In addition to computing the normative science performance metrics on the LSST data products, \faro provides a flexible and easy-to-use framework in which anyone can define and write a metric that will be computed on the LSST data products and run as part of the LSST science pipelines. 
The system is successfully running as part of nightly and weekly builds on precursor datasets to track the performance of the LSST science pipelines and provide regression analysis as development continues. 
\faro has also been successfully demonstrated at scale with the reprocessing of the DESC DC2 dataset \cite{2021ApJS..253...31L} at the Rubin Interim Data Facility (IDF) running on Google cloud \cite{2021arXiv211115030O} in the context of Rubin Data Preview 0.2 (DP0.2)\cite{RTN-001}. 
As we enter LSST commissioning, \faro will be a powerful tool to gain rapid insight into the quality of the data coming off the telescope.

\acknowledgments

We thank Yusra AlSayyad for her review of this manuscript.
This material is based upon work supported in part by the National Science Foundation through Cooperative Agreement AST-1258333 and Cooperative Support Agreement AST-1202910 managed by the Association of Universities for Research in Astronomy (AURA), and the Department of Energy under Contract No. DE-AC02-76SF00515 with the SLAC National Accelerator Laboratory managed by Stanford University. 
Additional Rubin Observatory funding comes from private donations, grants to universities, and in-kind support from LSSTC Institutional Members.

\bibliographystyle{spiebib}
\bibliography{local,lsst,lsst-dm,refs_ads,refs,books}

\end{document}